\documentclass[reprint,nofootinbib,superscriptaddress,showpacs]{revtex4-1}
\usepackage{amsmath}
\usepackage{amssymb}
\usepackage{dcolumn}
\usepackage{graphicx}
\usepackage{verbatim}
\usepackage{subfigure}
\usepackage[colorlinks, linkcolor=blue, citecolor=blue, urlcolor=blue]{hyperref}

\begin{document}
\title{\boldmath Improved DHOST Genesis}	
	
\author{Mian Zhu}
\email{mzhuan@connect.ust.hk}
\affiliation{Department of Physics, The Hong Kong University of Science and Technology, Clear Water Bay, Hong Kong S.A.R., P.R.China}
\affiliation{HKUST Jockey Club Institute for Advanced Study, The Hong Kong University of Science and Technology, Clear Water Bay, Hong Kong S.A.R., P.R.China}
\affiliation{Department of Astronomy, School of Physical Sciences, University of Science and Technology of China, Hefei, Anhui 230026, China}

\author{Yunlong Zheng}
\email{Corresponding author: zhyunl@ustc.edu.cn}
\affiliation{Department of Astronomy, School of Physical Sciences, University of Science and Technology of China, Hefei, Anhui 230026, China}
\affiliation{CAS Key Laboratory for Researches in Galaxies and Cosmology, University of Science and Technology of China, Hefei, Anhui 230026, China}
\affiliation{School of Astronomy and Space Science, University of Science and Technology of China, Hefei, Anhui 230026, China}
\affiliation{ICRANet, Piazza della Repubblica 10, I-65122 Pescara, Italy}

\begin{abstract}
We improve the DHOST Genesis proposed in \cite{Ilyas:2020zcb}, such that the near scale invariant scalar power spectrum can be generated from the model itself, without invoking extra mechanism like a string gas. Besides, the superluminality problem of scalar perturbation plagued in \cite{Ilyas:2020zcb} can be rescued by choosing proper DHOST action.
\end{abstract}

\maketitle
\flushbottom

\section{Introduction}

Currently, Inflation \cite{Guth:1980zm,Linde:1981mu,Albrecht:1982wi, Hawking:1981fz} is the standard paradigm of the very early universe. Alternative scenarios, like bounce cosmology \cite{Brandenberger:1992sy,Brandenberger:1993ef,Cai:2008qw,Cai:2012va, Yoshida:2017swb,Novello:2008ra,Lehners:2008vx,Cai:2014bea, Battefeld:2014uga,Brandenberger:2016vhg,Cai:2016hea}, emergent universe scenario \cite{Ellis:2002we,Ellis:2003qz} and string gas cosmology \cite{Brandenberger:1988aj,Nayeri:2005ck, Brandenberger:2006xi,Brandenberger:2006vv,He:2016uiy,Battefeld:2005av, Brandenberger:2011et,Brandenberger:2015kga}, are also of people's interests. It is believed that certain alternative scenarios can not only explain the formation of the Large Scale Structure of our universe as good as inflation, but also evade the initial spacetime singularity which plagues in inflationary cosmology \cite{Borde:1993xh,Borde:2001nh} (see \cite{Brandenberger:2009jq, Brandenberger:2011gk} for reviews on alternative cosmologies).

In this paper we focus on the emergent universe scenario, in which the universe is emergent from a quasi-Minkowskian spacetime. Such a scenario requires a violation of Null Energy Condition (NEC). One possible way to realize the NEC violation is introducing Horndeski/Generalized Galileon theory \cite{Nicolis:2008in, Deffayet:2011gz,Kobayashi:2011nu,Horndeski:1974wa}, which results in the Galileon Genesis model \cite{Creminelli:2010ba,Creminelli:2006xe} (see also earlier in \cite{Piao:2003ty}). The physics of Galileon Genesis is studied in \cite{LevasseurPerreault:2011mw,Wang:2012bq, Easson:2013bda,Rubakov:2013kaa,Elder:2013gya} and extended in \cite{Creminelli:2012my,Hinterbichler:2012fr,Hinterbichler:2012yn,Pirtskhalava:2014esa,Nishi:2015pta,Nishi:2016ljg,Cai:2017tku,Mironov:2019qjt,Ageeva:2020buc,Cai:2020qpu}.

The Galileon Genesis faces several challanges. Firstly, there are no-go theorems stating that no classically stable Genesis solution exists within the framework of Horndeski/Galileon theory \cite{Libanov:2016kfc,Kobayashi:2016xpl,Kolevatov:2016ppi,Akama:2017jsa}\footnote{It is proved in the framework of effective field theory \cite{Cai:2016thi,Cai:2017dyi,Cai:2017tku,Creminelli:2016zwa} that, a fully stable NEC violation can be realized only in theories beyond Horndeski.}. One possible ways out is to consider models which are strongly coupled in the asymptotic past/future \cite{Ijjas:2016vtq, Nishi:2016ljg}, at the cost of potentially triggering the strong-coupling problem \cite{Ageeva:2018lko,Ageeva:2020gti}. Besides, it is not very clear how such a Genesis scenario can be connected to the standard cosmology \footnote{In \cite{Kobayashi:2015gga}, a model of inflation preceded by (generalized) Galilean Genesis is proposed.}.

To evade the possible puzzles, we recently propose an emergent universe model from the DHOST cosmology \cite{Ilyas:2020zcb}. We modified the original action of Galileon Genesis such that the universe can gracefully exit from the quasi-Minkowskian state to a radiation dominated phase. Moreover, we introduce a DHOST coupling to cure the gradient instability. However, in our model the scale invariant scalar power spectrum cannot be generated within the model. To be consistent with astrophysical observations, we introduce a quasi-static Hagedorn phase from string gas cosmology, and hence the model itself is incomplete.

We complete the previous model by a slight deformation of the action. In Galileon Genesis, the behavior of Hubble parameter $H \propto (-t)^{-2\alpha - 1}$ determines the scalar spectrum index as $n_s = 5 - 2\alpha$ \cite{Nishi:2015pta, Nishi:2016ljg,Piao:2003ty,Liu:2011ns,Liu:2012ww}. We improve the action of our previous model \cite{Ilyas:2020zcb} accordingly, and show that the improved Genesis model can produce a scale invariant power spectrum. Hence our improved model can serve as a possible candidate for the early universe, without invoking additional mechanisms.

The paper is organized as follows. In section \ref{sec:model}, we briefly introduce the improved model and the genesis solution. The background dynamics is analyzed in section \ref{sec:bg}, where we verify that our model can gracefully exit to a radiation dominated universe. We work out the evolution of the scalar perturbation in section \ref{sec:scalarpt}, and show that the scale invariance can be obtained. The tensor perturbation is discussed in section \ref{sec:tensorpt}. Finally, we conclude our work in section \ref{sec:conclusion}.

Throughout the paper, we take the sign of the metric as (+,-,-,-). The canonical kinetic term of the scalar field $\pi$ is defined as $X \equiv \frac{1}{2}\nabla_{\mu}\pi \nabla^{\mu} \pi$. The d'Alembert operator is $\Box \equiv \nabla_{\mu}\nabla^{\mu}$. The reduced Planck mass $M_p^2 \equiv \frac{1}{8\pi G}$ is set to be 1.

\section{Improved DHOST Genesis}
\label{sec:model}
\subsection{Action}
The action of the model can be written as
\begin{equation}
	\label{eq:action}	
	S = \int d^4x \sqrt{-g} \left( \frac{R}{2} + \mathcal{L}_{H2} + \mathcal{L}_{H3} + \mathcal{L}_D \right) ~,
\end{equation}
where $\mathcal{L}_{H2}$ and $\mathcal{L}_{H3}$ come from the first two classes of Horndeski/Galileon theory \cite{Horndeski:1974wa,Deffayet:2011gz}, and $\mathcal{L}_D$ represents the type $^{(2)}N-II$ DHOST Lagrangian \cite{Langlois:2015skt,BenAchour:2016fzp,Langlois:2017mxy} \footnote{More explicitly, the action \eqref{eq:DHOST} belongs to the GLPV theory \cite{Gleyzes:2013ooa,Gleyzes:2014dya,Gleyzes:2014qga}, which propogates only one scalar degree of freedom \cite{Lin:2014jga,Gao:2014fra,Deffayet:2015qwa} and hence included by the DHOST theory.}:
\begin{equation}
	\mathcal{L}_{H2} = K(\pi,X) = - g_1(\pi) X + g_2(\pi) X^2 ~,
\end{equation}
\begin{equation}
	\mathcal{L}_{H3} = G(X)\Box X ~,~ G(X) = \gamma X^2 ~, 
\end{equation}
\begin{align}
	\label{eq:DHOST}
	\mathcal{L}_D & =  \frac{R}{2}h - \frac{h}{4X} \left[ \pi_{\mu \nu} \pi^{\mu \nu} - (\Box \pi)^2 \right] \nonumber \\
	& + \frac{h-2Xh_X}{4X^2} \left[ \pi_{\mu}\pi^{\mu \rho}\pi_{\rho \nu}\pi^{\nu} - (\Box \pi)\pi^{\mu}\pi_{\mu \nu}\pi^{\nu} \right] ~.
\end{align}
In equation \eqref{eq:action}, $R/2$ is the standard Einstein-Hilbert action. $g_1(\pi)$, $g_2(\pi)$ are funtion of the scalar field $\pi$ , while $h \equiv h(X)$ is a function of the canonical kinetic term $X$ only. Their detailed expressions are taken to be
\begin{equation}
	\label{eq:g12}
	g_1(\pi) = \frac{3}{2}f^2 e^{4\pi} \frac{1+e^{2\pi}}{1 + e^{m \pi}} ~,~ g_2(\pi) = e^{2\pi} \frac{1 + e^{2\pi}}{1+e^{4\pi}} ~, 
\end{equation}
\begin{equation}
	h(X) = d_1 X + d_2 X^2 ~,
\end{equation}
in which $m>6$ is a fixed parameter which determines the behavior when the universe exiting the quasi-Minkowskian state. As we will show in next sections, the value of $m$ doesn't change the scale dependence, so we will simply take $m=7$ through out this paper.


From equation \eqref{eq:g12}, we have
\begin{equation}
\label{eq:glimit-inf}	
	\lim_{\pi \to -\infty} g_1(\pi) = \frac{3}{2}f^2e^{4\pi} ~,~ \lim_{\pi \to -\infty} g_2(\pi) = e^{2\pi} ~; 
\end{equation}
\begin{equation}
\label{eq:glimit+inf}	
	\lim_{\pi \to \infty} g_1(\pi) = 0 ~,~ \lim_{\pi \to \infty} g_2(\pi) = 1 ~.
\end{equation}

Applying the FLRW background to the action \eqref{eq:action}, we recover the Friedmann equations 
\begin{equation}
\label{eq:Friedmann}	
	3H^2 = \rho_{\pi} ~,~ -2\dot{H} = \rho_{\pi} + p_{\pi} ~,
\end{equation}
where $H$ is the Hubble parameter, $\rho_{\pi}$ and $p_{\pi}$ are the density and pressure of the matter field $\pi$, expressed by
\begin{equation}
	\label{eq:rho}
	\rho_{\pi} = -\frac{1}{2} g_1 \dot{\pi}^2 + \frac{3}{4}g_2 \dot{\pi}^4 + 3H\gamma \dot{\pi}^5 ~,
\end{equation}
\begin{equation}
	\label{eq:p}
	p_{\pi} = -\frac{1}{2}g_1 \dot{\pi}^2 + \frac{1}{4}g_2 \dot{\pi}^4 - \gamma \dot{\pi}^4\ddot{\pi} ~.
\end{equation}

We see from equation \eqref{eq:rho} and \eqref{eq:p} that the DHOST term \eqref{eq:DHOST} doesn't enter into the background dynamics. This fact is already found in \cite{Ilyas:2020qja,Ilyas:2020zcb} as long as the DHOST coupling is non-trivial only during the NEC violation phase. Moreover, observational singnals from scalar perturbation, such as the scalar spectra index and tensor-to-scalar ratio, will not be affected by DHOST action in bounce cosmology \cite{Zhu:2021whu}, and in section \ref{sec:scalarpt}, we will similarly find that the scalar spectra index is independent of the DHOST action.

\subsection{The Genesis Solution}
\label{sec:bggenesis}
We assume the initial condition of scalar field $\pi$ at far past to be $\pi \ll -1$. From equation \eqref{eq:glimit-inf}, we see the first Friedmann's equation becomes
\begin{equation}
	3H^2 = -\frac{3}{4} f^2 e^{4\pi} \dot{\pi}^2 + \frac{3}{4}e^{2\pi} \dot{\pi}^4 + 3H\gamma \dot{\pi}^5 ~,
\end{equation}
and there is a quasi-Minkowskian solution with $H \approx 0$ as long as 
\begin{equation}
\label{eq:piinfty}	
	\dot{\pi}^2 e^{-2\pi} = f^2 ~\to~ \pi = \ln \left[ \frac{1}{f(-t)} \right] ~.
\end{equation}

Substituting equation \eqref{eq:piinfty} into equation \eqref{eq:p}, the pressure becomes
\begin{equation}
	p_{\pi} = -\frac{1}{(-t)^6} \left( \frac{1}{2f^2} + \gamma \right) ~,
\end{equation}
which is of order $\mathcal{O}(t^{-6})$ and larger than that of the density. So the universe will deviate from the quasi-Minkowskian configuration by the negative and non-vanishing pressure.

\subsection{Model Parameters}
To illustrate the dynamics of the universe and numerically verify the approximations used in next sections, we will use the following parameters through out this paper:
\begin{equation}
	\label{eq:parameter}
	f = 0.8 ~,~ \gamma = 1 ~,~ d_1 = 0 ~,~ d_2 = 6 ~.
\end{equation}

As we shall see in section \ref{sec:spt}, the sound speed of scalar perturbation in pre-emergent period is $c_{s0}^2 = (4f^2 \gamma - 1)/3$. To avoid the gradient instability and superluminality problem in this period, we have $0 < c_{s0}^2 < 1$ and the model parameters are constrained by 
\begin{equation}
	\frac{1}{4} < \gamma f^2 < 1 ~.
\end{equation}
Finally, we will show in section \ref{sec:stability} that the DHOST coupling chosen in equation \eqref{eq:parameter} can evade the gradient instability and superluminality problem of scalar perturbation in the whole cosmological evolution.

\section{Background Evolution}
\label{sec:bg}
In this section we briefly discuss the background dynamics of our model. As shown in section \ref{sec:bggenesis}, at far past $t \to -\infty$, the model permits a quasi-Minkowskian solution described by equation \eqref{eq:piinfty}. Substituting the solution \eqref{eq:piinfty} into the second Friedmann  equation \eqref{eq:Friedmann} and using $p_{\pi} \gg \rho_{\pi}$, we have 
\begin{equation}
	-2\dot{H} \simeq p_{\pi} = -\frac{1}{(-t)^6} \left( \frac{1}{2f^2} + \gamma \right) ~,
\end{equation}
so that
\begin{equation}
	\label{eq:Ht<0}
    H = \frac{h_0}{(-t)^5} ~,~ h_0 \equiv \frac{1}{10} \left( \frac{1}{2f^2} + \gamma \right) ~.
\end{equation}

When the universe evolves towards the emergent event $t=0$, the approximated equations \eqref{eq:piinfty}, \eqref{eq:Ht<0} become invalid, and the universe will deviate from the quasi-Minkowskian state. We define a typical time scale $t_E$, so that the approximations \eqref{eq:piinfty} and \eqref{eq:Ht<0} is valid when $t < t_E < 0$. For the specific set of parameters \eqref{eq:parameter}, $t_E $ is of order $\mathcal{O}(1)$. After the emergent event $t=0$, the scalar field $\pi$ become positive and the function $g_1(\pi)$, $g_2(\pi)$ quickly turns to the form \eqref{eq:glimit+inf}. The pressure and density then becomes
\begin{equation}
	\rho_{\pi} = \frac{3}{4}\dot{\pi}^4 + 3H\gamma \dot{\pi}^5 ~,~ p_{\pi} = \frac{1}{4}\dot{\pi}^4 - \gamma \dot{\pi}^4 \ddot{\pi} ~,
\end{equation}
and there is an approximated solution
\begin{equation}
	a \propto t^{\frac{1}{2}} ~,~ \dot{\pi} \propto t^{-\frac{1}{2}} ~,
\end{equation}
which represents a radiation dominated universe.

\begin{figure}[t]
	\centering
	\includegraphics[width=.23\textwidth]{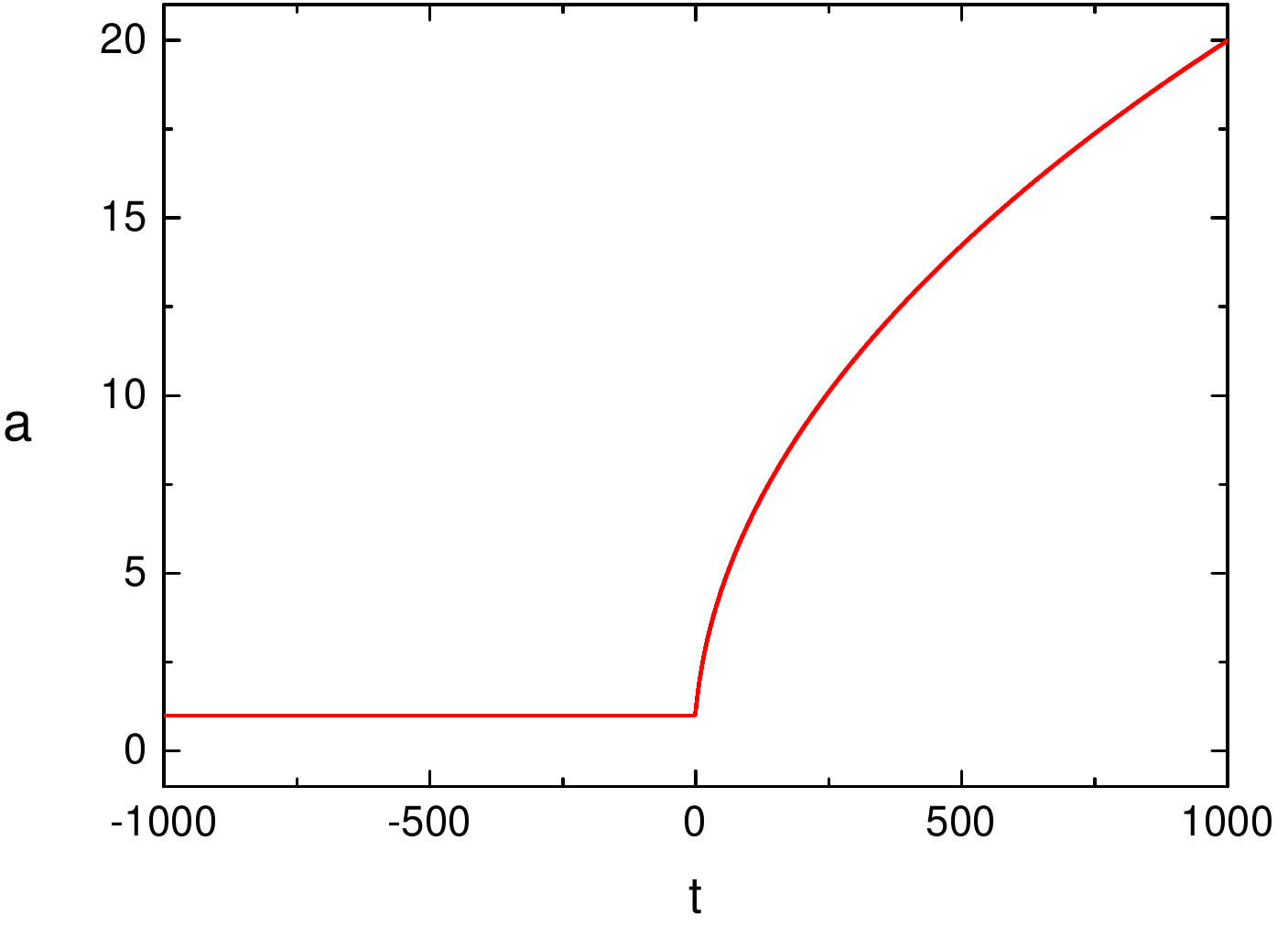} \! 
	\includegraphics[width=.23\textwidth]{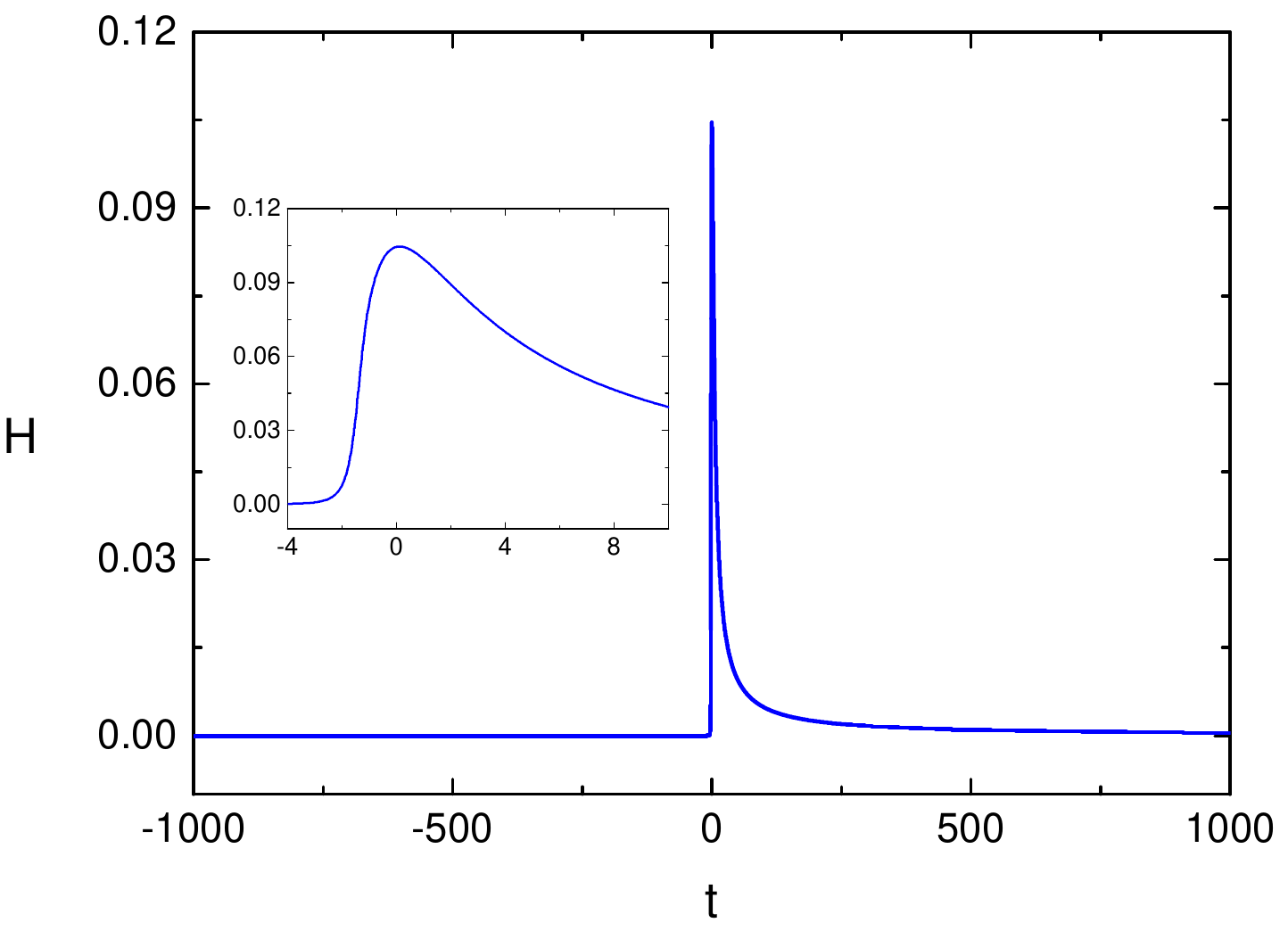} \\
	\includegraphics[width=.23\textwidth]{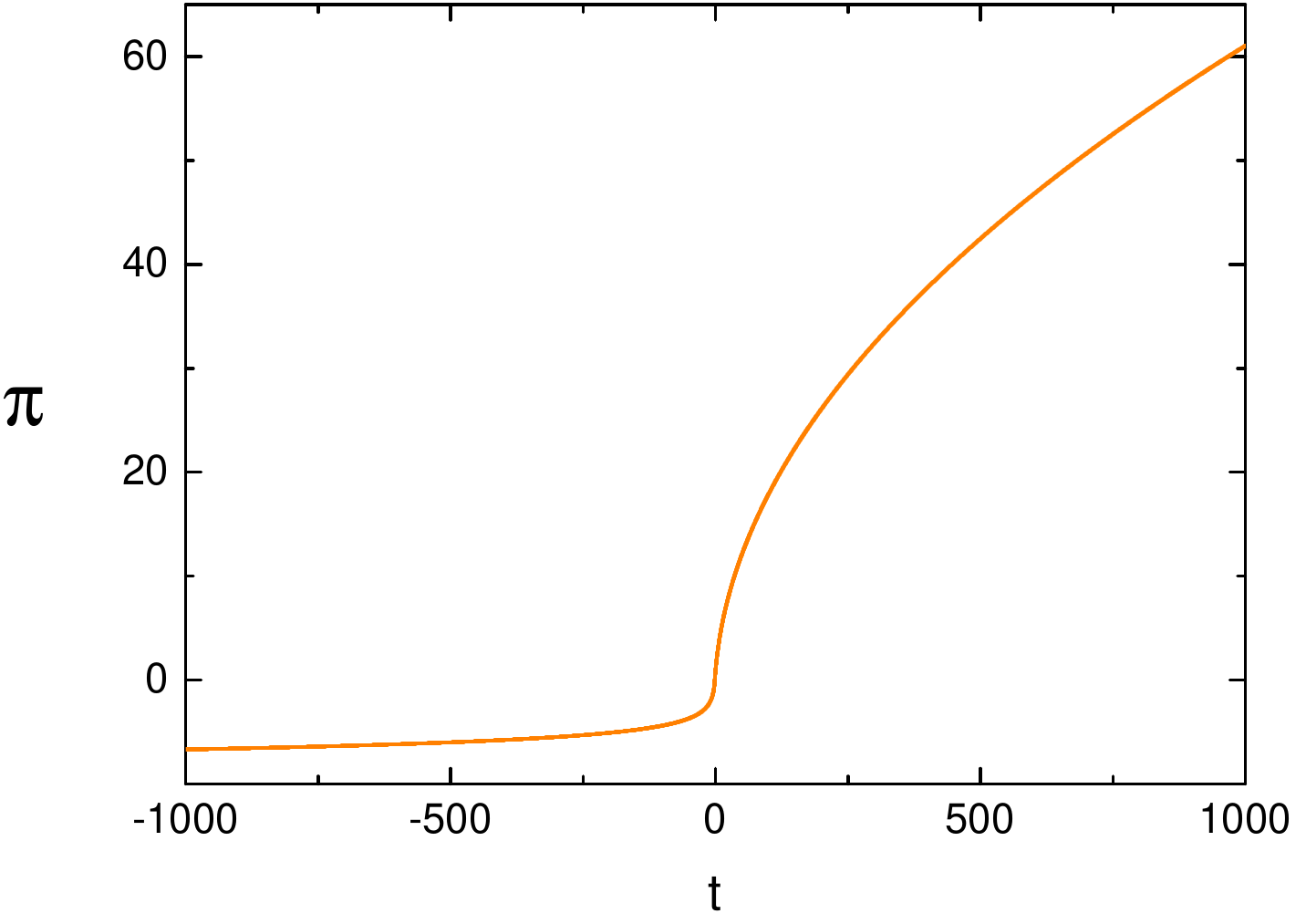} \!
	\includegraphics[width=.23\textwidth]{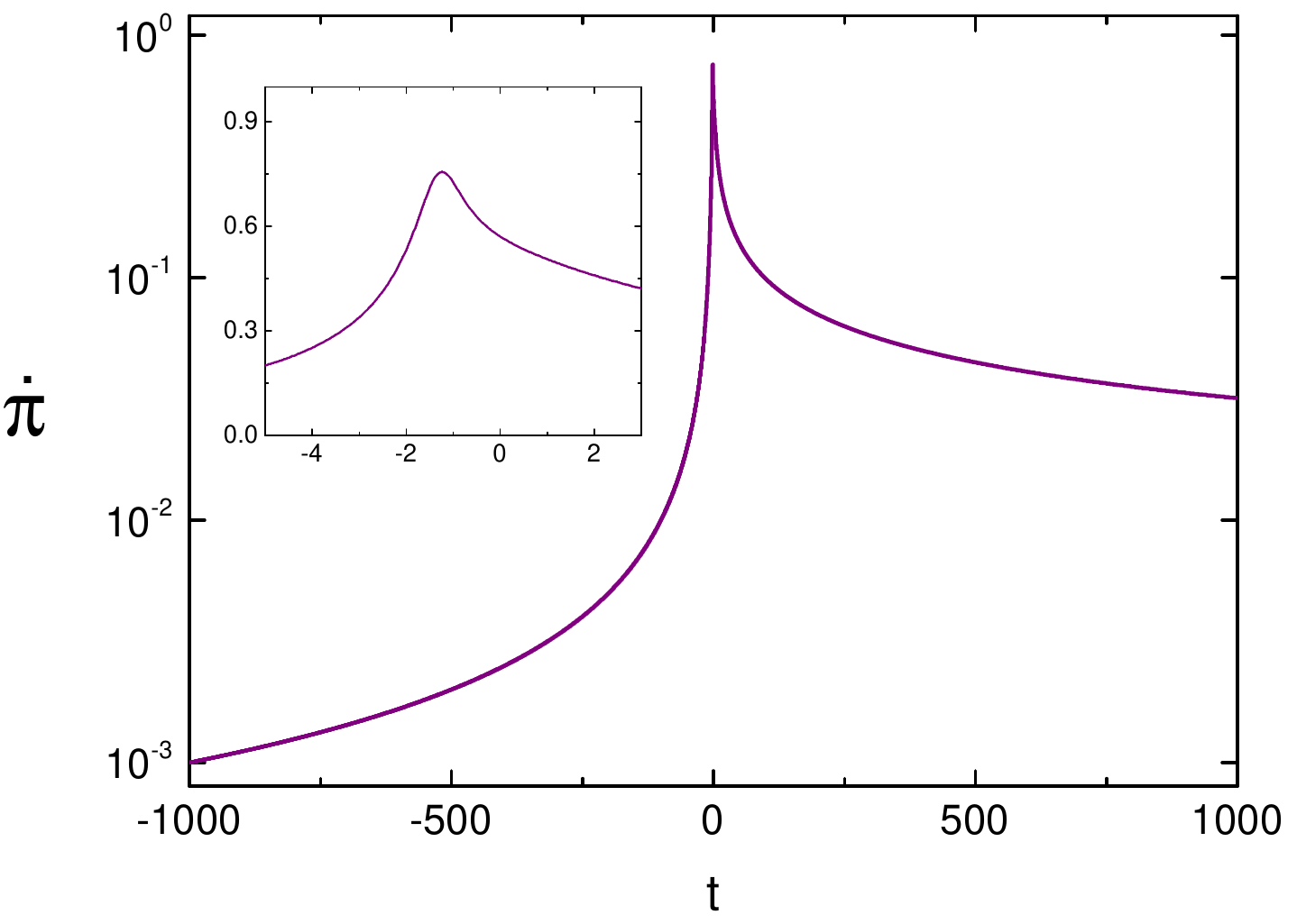}
	\caption{Dynamics of background geometry $a$, $H$ and scalar field $\pi$, $\dot{\pi}$ as a function of cosmic time $t$ with the special parameter set \eqref{eq:parameter}.}
	\label{fig:bg}
\end{figure}

As shown above, the universe starts from a quasi-Minkowskian state, then quit the status around the emergent event $t = 0$, and finally transits to a radiation dominated universe, thus connected to the standard Cosmology. Our model then provides a graceful exit mechanism to the regular Galileon Genesis scenario. We illustrate the cosmological evolution numerically in figure \ref{fig:bg}.

\section{Scalar Perturbation}
\label{sec:scalarpt}

\subsection{Dynamic equation and the stability issue}
\label{sec:stability}
In this section, we study the scalar perturbation of the current model. Applying the previous result developed in \cite{Ilyas:2020qja,Ilyas:2020zcb}, the scalar perturbation at linear level is expressed by 
\begin{equation}
\label{eq:S2s}	
	S_{2,s} = \int d\tau d^3x \frac{z_s^2}{2} \left[ \zeta^{\prime 2} - c_s^2 (\partial_i \zeta)^2 \right] ~,
\end{equation}
where $\zeta$ is the curvature perturbation and
\begin{equation}
	\frac{z_s^2}{2a^2} = 3 + 2\frac{\dot{\pi}^2(-g_1 + 3g_2\dot{\pi}^2) + 18H\gamma \dot{\pi}^5 - 6H^2}{(\gamma \dot{\pi}^5 - 2H)^2} ~,
\end{equation}
\begin{equation}
	\left( - \frac{z_s^2}{2a^2} \right) c_s^2 = 1 + h + \frac{2}{a} \frac{d}{dt} \frac{a \left( h_X \dot{\pi}^2 - h - 1 \right)}{2H - \gamma \dot{\pi}^5} ~.
\end{equation}

The dynamical equation of the scalar perturbation is the standard Mukhanov-Sasaki(MS) equation \cite{Sasaki:1983kd,Kodama:1985bj, Mukhanov:1988jd}:
\begin{equation}
	\label{eq:MSscalar}
	v_k^{\prime \prime} + \left(k^2c_s^2 - \frac{z_s^{\prime \prime}}{z_s}\right) v_k = 0 ~,
\end{equation}
where $v_k \equiv z_s \zeta$ is the MS variable, and a prime denotes differentiation with respect to the conformal time $d\eta = dt/a$. 

\begin{figure}[t]
    \includegraphics[width=.21\textwidth]{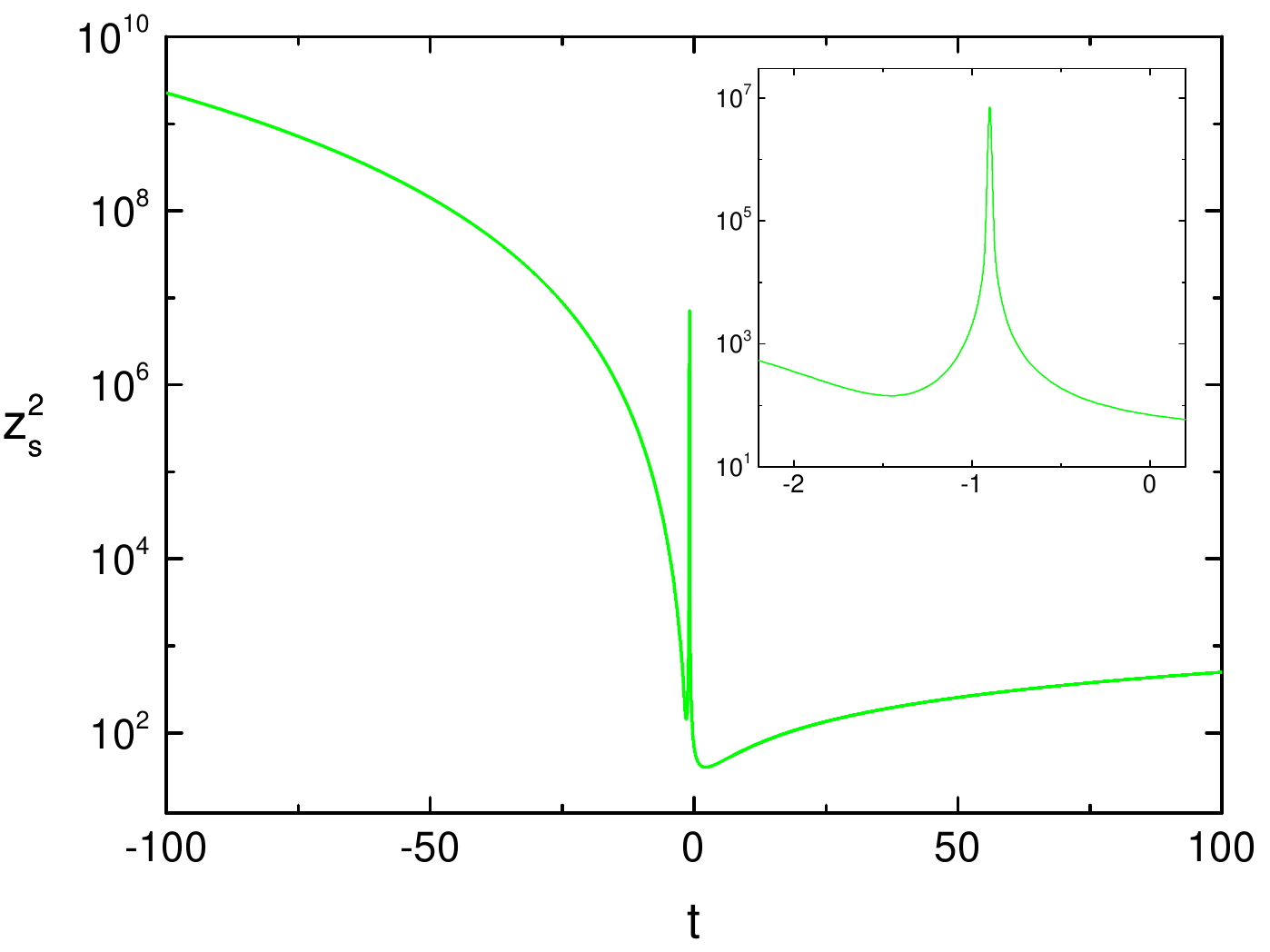} \! 
    \includegraphics[width=.2\textwidth]{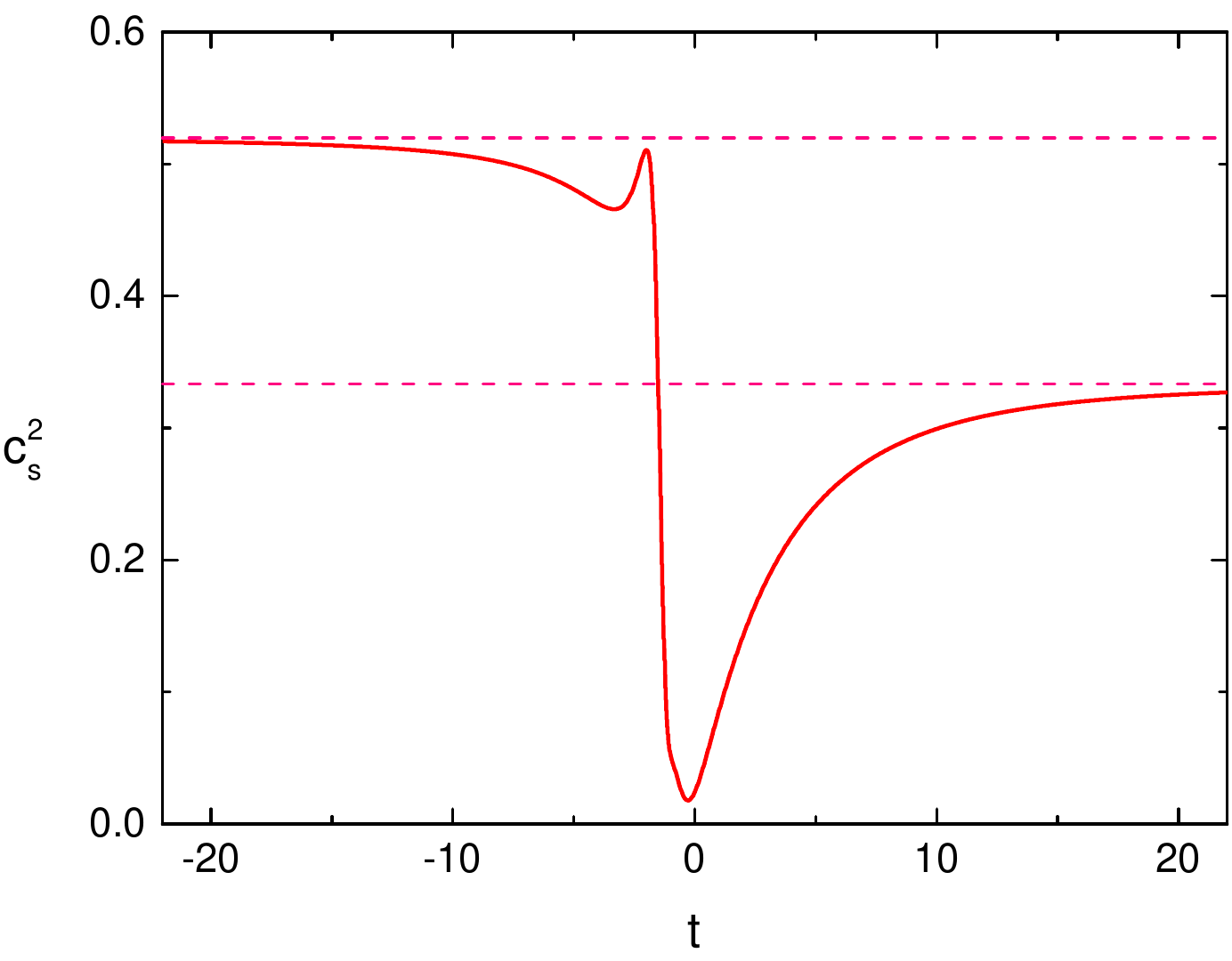} 
    \caption{Dynamics of $z_s^2$ and $c_s^2$ with respect to cosmic time $t$. The model parameters are set by \eqref{eq:parameter}.}
    \label{fig:scalarpt}
\end{figure}

To obtain a healthy model, we shall avoid the gradient instability and the existence   of the ghost. This requires 
\begin{equation}
	z_s^2 > 0 ~,~ c_s^2 > 0 ~. 
\end{equation}
We numerically illustrate the dynamics of $z_s^2$ and $c_s^2$ in figure \ref{fig:scalarpt} with parameters being \eqref{eq:parameter}. The gradient and ghost instabilities are absent. Moreover, the scalar sound speed does not exceed the speed of light, i.e. $c_s^2 < 1$, and hence the superluminality problem plagued in \cite{Ilyas:2020zcb} is resolved.

\subsection{General expression for the scalar power spectrum}
\label{sec:spt}
In the pre-emergent period $t<0$, the universe is almost static and $d\eta = dt/a \simeq dt$. By properly setting the integration constant, we can interchangably use $\eta$ and $t$ in this region. Moreover, the DHOST coupling $h \approx 0$, along with the asymptotic behavior \eqref{eq:glimit-inf} and \eqref{eq:Ht<0}, the parameters reduce to
\begin{equation}
	\label{eq:zscst<0}
	z_s^2 = \frac{600f^2}{\left(1-8f^2\gamma \right)^2}(-t)^4 ~,~ c_s^2 = \frac{4f^2\gamma-1}{3} \equiv c_{s0}^2 ~.
\end{equation}

In equation \eqref{eq:zscst<0}, only the highest order of $(-t)$ are kept. The MS equation \eqref{eq:MSscalar} becomes 
\begin{equation}
	v_k^{\prime \prime} + \left(k^2c_{s0}^2 - \frac{2}{\eta^2}\right)v_k = 0 ~,
\end{equation}
whose general solution is
\begin{equation}
	\label{eq:vketa}
	v_k(\eta) \propto \sqrt{k \eta} \left[ c_1(k)J_{\frac{3}{2}}(kc_{s0} \eta) + c_2(k) Y_{\frac{3}{2}} (kc_{s0} \eta) \right] ~,
\end{equation}
where $J_a(z)$ and $Y_a(z)$ are the Bessel functions. 


Before proceeding, we mention that in generalized Galilean Genesis, the curvature perturbation $\zeta$ grows even on superhorizon scales when the second Horndeski term $\mathcal{L}_{H3} \propto X^{\alpha} \Box \pi$ satisfies $\alpha > 1/2$ \cite{Creminelli:2010ba, Nishi:2016ljg}. Our model corresponds to the $\alpha = 2$ case, so we don't need to determine the time for each wavemode to exit the horizon during the pre-emergent period. However, after the emergent event $t=0$, the universe becomes radiation dominated and the scalar perturbation contains a constant mode and a decreasing mode. Hence the dominated mode which is superhorizon at $t>0$ will be the constant one, and the scalar perturbation will then remains invariant, just as the case in standard cosmology.

Now, as we argued in \cite{Ilyas:2020zcb}, the wavemodes of observational interests will all become superhorizon in the pre-emergent period $t<0$. So we should evaluate the dynamics of scalar perturbation from past infinity to the critical time scale $\eta_E = t_E$, where the emergent solution breaks down. After $t=t_E$, the universe quickly turns to a radiation dominated period, and the scalar perturbation becomes conserved. 

Hence, the scalar power spectrum can be evaluated by
\begin{equation}
\label{eq:Pzeta}	
	P_{\zeta}(k) = \frac{k^3}{2\pi^2} \frac{|v_k(\eta_E)|^2}{z_s(\eta_E)^2} ~.
\end{equation}

\subsection{Vacuum initial condition}
In this section we impose the vacuum initial condition and work out the scalar spectra index. Firstly, the argument of the Bessel function $kc_{s0}\eta_E \ll 1$, so we can expand the Bessel function with small argument approximation: 
\begin{equation}
	\label{eq:Besselt=0}
	J_a(z) \sim \frac{1}{\Gamma(a+1)} \left( \frac{z}{2} \right)^a ~,~ Y_a(z) \sim - \frac{\Gamma(a)}{\pi} \left( \frac{2}{z} \right)^a ~,
\end{equation}
and the Bessel $Y$ function will dominate at $\eta = \eta_E$. The MS variable is approximately
\begin{equation}
	\label{eq:vketaE}
	v_k(\eta_H) \propto c_2(k) \sqrt{k\eta_H} (k\eta_H)^{-\frac{3}{2}} = c_2(k)(k\eta_H)^{-1} ~.
\end{equation}

Now we determine the $c_2(k)$ term from the vacuum initial condition. In the far past, the linearized action \eqref{eq:S2s} is of a harmonic oscillator form with a nearly constant sound speed $c_{s0}$. So we can impose the quantum vacuum initial condition
\begin{equation}
	\label{eq:vkinitial}
	v_k(\eta) \to \frac{e^{-i c_{s0} k (\eta - \eta_0)}}{\sqrt{2k}} ~, 
\end{equation}
where $\eta_0$ is an integration constant representing the initial phase. Comparing equation \eqref{eq:vkinitial} and \eqref{eq:vketa}, and with the help of large argument expansion of Bessel function, we find 
\begin{equation}
	\label{eq:c2k}
	c_2(k) \propto k^{-1/2} ~.
\end{equation}

Substituting equation \eqref{eq:c2k} and \eqref{eq:vketaE} into the expression of scalar spectrum \eqref{eq:Pzeta}, we find
\begin{equation}
	P_{\zeta} \propto k^3 \times |k^{-\frac{3}{2}}|^2 = k^0 ~,
\end{equation}
and the scale invariant scalar power spectrum is acquired. 

\section{Tensor Perturbation}
\label{sec:tensorpt}
The tensor perturbation at the linearized level is \cite{Ilyas:2020zcb}
\begin{equation}
	S_{2,T} = \int d\eta d^3x \frac{a^2}{8} \left[ \gamma_{ij}^{\prime 2} - (1+h) \gamma_{ij}^2 \right] ~,
\end{equation}
in which $\gamma_{ij}$ represents the tensor perturbation.

\begin{figure}[t]
	\includegraphics[width=.3\textwidth]{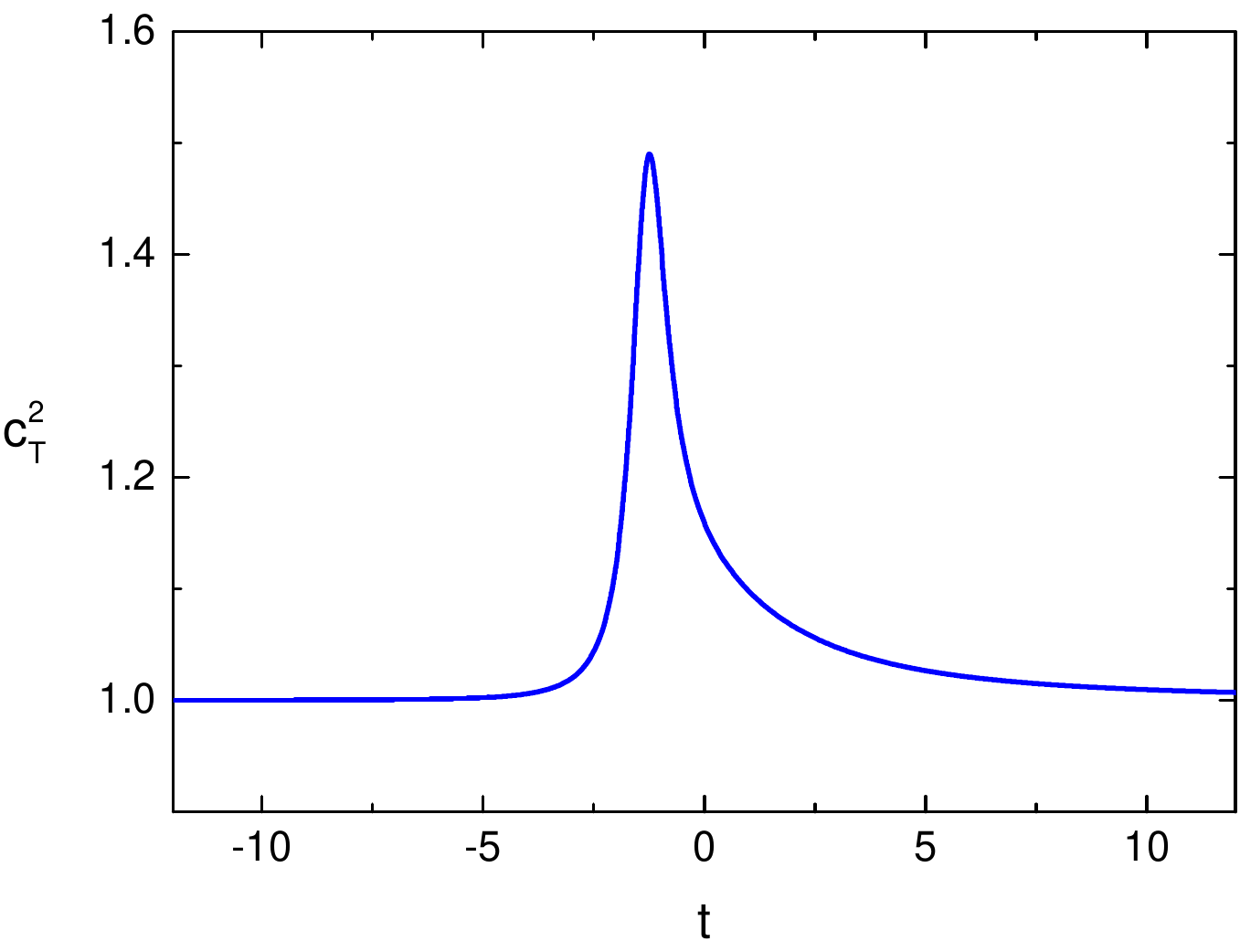}
	\caption{Dynamics of $c_T^2$ with respect to cosmic time $t$. The model parameters are set by \eqref{eq:parameter}.}
	\label{fig:ct2}
\end{figure}

Since the coefficient of the kinetic term $\gamma_{ij}^{\prime 2}$ is always positive, there is no ghost model of tensor perturbation. We numerically plot the sound speed of tensor perturbation $c_T^2 = 1+h$ in figure \ref{fig:ct2}, and show that there is no gradient instability as well.

The tensor perturbation contains two modes. In our case the two polarization modes propagate at the same speed \cite{Gao:2019liu}, so we can collectively denote them as $\gamma$, and the MS equation becomes
\begin{equation}
	\label{eq:MStensor}
	\nu_k^{\prime \prime} + \left(c_T^2k^2 - \frac{a^{\prime \prime}}{a} \right) \nu_k = 0 ~,
\end{equation}
where $\nu_k \equiv 1/2 a\gamma_k$ is the MS variable for tensor perturbation.

In the radiation dominated period, $a \propto t^{1/2}$, so the conformal time $\eta = \int dt/a \propto t^{1/2}$. We then have $a \propto \eta$ and $a^{\prime \prime}/a$ vanishes. Hence the tensor perturbation propagates as a plane wave, and the tensor spectrum index is invariant in this period. Similarly, in the pre-emergent period the background is quasi-Minkowskian, and the same conclusion applied. One may also see the fact by noticing that, $H \propto (-t)^{-5}$ when $t<t_E$ and hence $a^{\prime \prime}/a \propto (-t)^{-6}$, which is negligible compared to $c_T^2k^2$ in the MS equation \eqref{eq:MStensor}. Finally, the duration of the transition period is small (of order $\mathcal{O}(10^0)$ in our set of parameters). Since there is no divergent part in MS equation \eqref{eq:MStensor} during this period, the contribution from this period to tensor perturbation will be suppressed by its short duration, so the tensor spectrum index is also invariant. We then conclude that, the tensor spectrum index we observed is just that in the beginning of the universe. Substituting the vacuum initial condition $\nu_k \propto k^{-1/2}$ \footnote{The expression is analog to that for scalar perturbation.} into the expression for tensor spectrum $P_t \propto k^3 |\nu_k|^2$, we see that $n_t = \frac{d\ln P_t}{d\ln k} = 2$, which is consistent with previous studies \cite{Creminelli:2010ba,Nishi:2015pta}.

\section{Conclusion and Outlook}
\label{sec:conclusion}
In this paper we improve the DHOST Genesis model \cite{Ilyas:2020zcb} to get the scale invariance scalar power spectrum. Compared to the original Galileon Genesis model \cite{Creminelli:2010ba}, the improved model consists three new ingredient. The deformed Horndeski term $\mathcal{L}_{H2}$ provides the realization for a graceful exit from emergent universe to standard cosmology; the $\mathcal{L}_{H3}$ term determines the scalar spectrum index, resulting in the scale invariance; The DHOST term $\mathcal{L}_D$ modifies the sound speed of scalar perturbation $c_s^2$ near the emergent event $t=0$ without altering the background evolution, hence solves the gradient instability. The emergent universe scenario presented here is then healthy and can meet the basic observational constraint without other mechanism invoked.

There are still plenty of rooms for us to improve the current model. Firstly, most of the emergent universe models (including the one here) predict a blue tensor spectrum with $n_t = 2$ (counterexamples can be found in, say, \cite{Cai:2016gjd, Nishi:2016ljg}), so updating the current model to allow more possibility for $n_t$ would be worth investigating. Secondly, it is interesting to develop genesis models with alternative mechanism (like the curvaton mechanism \cite{Creminelli:2010ba, Wang:2012bq}) to realize the scale invariance, and see whether they can be distinguished by more observational data. Thirdly, it is worthy to develop the model to make the sound speed $c_T^2$ are less than $c^2 = 1$, hence evading the superluminality issue. Finally, a detailed phase space analyze can help to decide whether there is fine-tuning problem in our model.

\section*{ACKNOWLEDGMENTS}
We are grateful to Yifu Cai, Robert Brandenberger, Yong Cai, Yi Wang, Damien Easson, Zhibang Yao, Xian Gao, Emmanuel Saridakis, Misao Sasaki and Dong-Gang Wang  for stimulating discussions. 
This work is supported in part by the NSFC (Nos. 11847239, 11961131007,11421303), by the Fundamental Research Funds for Central Universities,  by the China Scholarship Council (CSC No.202006345019),  and by GRF Grant 16304418 from the Research Grants Council of Hong Kong.  YZ would like to thank the ICRANet for their hospitality during his visit.  
All numeric are operated on the computer clusters {\it LINDA \& JUDY} in the particle cosmology group at USTC.

\bibliography{reference}
\bibliographystyle{unsrt}

\end{document}